\documentclass[12pt]{article}
\usepackage{graphicx}
\usepackage{epstopdf, epsfig}
\pdfoutput=1

\usepackage[a4paper, margin=2.5cm]{geometry}
\usepackage{amsmath} 
\usepackage{multirow} 
\usepackage{amssymb}
\usepackage{natbib}
\usepackage{caption}
\usepackage{bm}
\usepackage{tabularx}
\usepackage{caption}
\usepackage{subcaption}

\usepackage{placeins} 
\usepackage{hyperref}
\usepackage[mathscr]{euscript}
 \let\mathscr\relax
\usepackage[scr]{rsfso}

\usepackage{enumitem}

\usepackage{cleveref}
\crefformat{section}{\S#2#1#3}

\title{Efficiency of turbulence}

\author{\small A. Lopez$^1$, A. Barral$^1$, G. Costa$^1$, Q. Pikeroen$^1$, V. Shukla$^2$, B\'ereng\`ere Dubrulle$^{1\dagger}$}

\date{\small $^1$Universit\'e Paris-Saclay, CEA, CNRS, SPEC, 91191, Gif-sur-Yvette, France\\
$^2$Department of Physics,
Indian Institute of Technology Kharagpur,
Kharagpur, India \\
$^\dagger$berengere.dubrulle@cea.fr
}

\DeclareMathOperator{\Ra}{Ra}
\DeclareMathOperator{\Nu}{Nu}

\def\kmax{k_{\rm max}}

\begin{document}


\maketitle


\begin{abstract}

We consider the efficiency of turbulence, a dimensionless parameter that characterises the fraction of the input energy stored into a turbulent flow field. We first show that the inverse of the efficiency provides an upper bound for the dimensionless energy injection in a turbulent flow. We analyse the efficiency of turbulence for different flows using numerical and experimental data. Our analysis suggests that efficiency is bounded from above, and, in some cases, saturates following a power law reminiscent of phase transitions and bifurcations. We show that for the von K\'arm\'an flow the efficiency saturation is insensitive to the details of the forcing impellers. In the case of Rayleigh-B\'enard convection, we show that within the Grossmann and Lohse model, the efficiency saturates in the inviscid limit, while the dimensionless kinetic energy injection/dissipation goes to zero. In the case of pipe flow, we show that saturation of the efficiency cannot be excluded, but would be incompatible with the Prandtl law of the drag friction coefficient. Furthermore, if the power law behaviour holds for the efficiency saturation, it can explain the kinetic energy and the energy dissipation defect laws proposed for the shear flows. Efficiency saturation is an interesting empirical property of turbulence that may help in evaluating the ``closeness" of experimental and numerical data to the true turbulent regime, wherein the kinetic energy saturates to its inviscid limit. 

\end{abstract}

\section{Introduction}

The Joule's experiment of 1845 showed that when water in a bucket is mechanically stirred, the temperature of the water rises. The experiment illustrates the conversion of mechanical energy into thermal energy in fluids in agreement with the first law of thermodynamics. Richardson cascade phenomenology suggests that this energy conversion is achieved via a two-step process: first, the creation of finer and finer energy-transporting structures, also called eddies, then when the size of the structures is reduced to the Kolmogorov length $\eta$, viscosity effectively transforms the mechanical energy into heat. However, in this process, some finite amount of mechanical energy is stored within the fluid through erratic movements of eddies of all sizes. 

The efficiency of this storage can be characterised by introducing a dimensionless number ${\cal{E}}=E/E_f$, where $E$ is the stored kinetic energy per unit mass and $E_f= F_0 L_f$ is the measure of the mechanical energy injected per unit mass due to an applied force of magnitude $F_0$ acting at length scale $L_f$. Despite its obvious interpretation, this number is seldom used to describe characteristics of turbulence; instead, we often use Reynolds number $Re=L_f \sqrt{2E}/\nu$ or the Grashof number $Gr=F_0 L^3_f/\nu$, where $\nu$ is the fluid viscosity. This allows us to express the efficiency number as ${\cal{E}}=Re^2/2Gr$, or it inverse, the inefficiency $1/{\cal{E}}$ ~\citep{eyink2024}. However, the independence of ${\cal{E}}$ with respect to the viscosity makes it suitable for exploring the asymptotic regimes of turbulence, wherein the viscosity does not play a role anymore. In this paper, we show an application of this idea to the problem of intensity of turbulence for injection and dissipation of energy within a flow and its connection to the empirical defect laws that have been recently proposed for turbulent flows near a wall in~\cite{Chen_Sreenivasan_2021}.

\section{Efficiency of turbulence}

\subsection{Definitions}

We consider the forced \textbf{incompressible Navier-Stokes equations} (INSE) given by
\begin{eqnarray}\label{NS}
\partial_t\bm{u} + (\bm{u}\cdot\bm{\nabla})\bm{u}&=& -\bm{\nabla}p + \nu\bm{\Delta}\bm{u} + \bm{f},\\
\nabla\cdot \bm{u}&=&0,
\end{eqnarray}
where $\bm{u}$ is the velocity field, $p$ is the pressure, $\bm{f}$ is the external force, $\nu$ is the viscosity and the constant density is set to $\rho=1$. We define the total kinetic energy $E=\langle \bm{u}\cdot \bm{u}\rangle/2$ and the forcing amplitude $F_0=\langle \bm{f}\cdot \bm{f}\rangle^{1/2}$, where $\langle \cdot \rangle$ denotes the spatial average. We further introduce $L_f$, a characteristic forcing length scale defined by 
\begin{equation}
\frac{(2\pi)^2}{L^2_f} = \frac{1}{3} \frac{\langle \nabla \bm{f}\cdot \nabla \bm{f} \rangle}{\langle \bm{f}\cdot \bm{f}\rangle}.
\end{equation}
We can then non-dimensionalise the Navier-Stokes equations (\ref{NS}) using $\sqrt{2E}$ as unit of velocity, $L_f$ as 
unit of length as:
\begin{eqnarray}\label{NSND}
\partial_t \tilde{\bm{u}} + (\tilde{\bm{u}}\cdot\bm{\nabla})\tilde{\bm{u}}&=& -\bm{\nabla}\tilde{p} + \frac{1}{Re}\bm{\Delta}\tilde{\bm{u}} +\frac{1}{2{\cal E}} \tilde{\bm{f}},\\
\nabla\cdot \tilde{\bm{u}}&=&0,
\end{eqnarray}
where $\tilde{\bm{u}}$ and $\tilde{\bm{f}}$ are variable of unit average norm and
\begin{eqnarray}
Re&=& \frac{\sqrt{2E} L_f}{\nu},\nonumber\\
{\cal{E}} &=& \frac{\langle \bm{u}\cdot \bm{u}\rangle}{2L_f\langle \bm{f}\cdot \bm{f}\rangle^{1/2}},
\label{Re&E}
\end{eqnarray}
are the Reynolds number and the efficiency. This non- dimensionalisation illustrates that the parameter space depends on both $Re$ and $\cal{E}$. In most application, one fixes $F_0$, $L_f$ and $\nu$ and measures $\langle E \rangle$.
This amounts to fixing the Grashof number $Gr=Re^2/{\cal E}$ as the external control parameter, and use $Re$ and ${\cal E}$ as diagnostic parameters. However, one could also fix either $Re$ or ${\cal E}$ as the control parameter, and measure
the other one as a diagnostic parameter, see below.
 
\subsection{Energy injection}
 The energy injected per unit mass $\epsilon$ is given by
\begin{equation}
\epsilon =\langle \bm{f}\cdot \bm{u} \rangle.
\end{equation}
In stationary turbulence, and in absence of any singularities, energy injection equals the energy dissipation on average. In many applications, we are interested in knowing the amount of work per unit volume $\epsilon$ that must be performed to maintain a fluid of viscosity $\nu$, stirred by an external force of magnitude $F_0$, in a stationary state characterised by a mean kinetic energy per unit mass $\langle E \rangle$. This will require us to study $\epsilon$ as function of $\nu$ and $\langle E \rangle$, or, in dimensionless form, $D_\epsilon=\frac{|\epsilon| L_f}{\sqrt{2}E^{3/2}}$ as a function of $Re$. In this study, we will also focus on another problem, namely, the amount of viscosity $\nu$ required to sustain a given mean mechanical energy per unit mass $\langle E \rangle$ in a fluid, when acted upon by a force of magnitude $F_0$. In dimensionless form, this corresponds to studying 
$1/Re$ as a function of ${\cal E}$.

The previous discussion holds in incompressible, homogeneous isotropic turbulence in a statistically stationary state. Wall bounded turbulence like pipe flow add additional boundary terms to the energy balance, but can still be approached using the efficiency prism, as we show below. Similarly, convection adds new energy source terms, but we show that it can benefit from the efficiency consideration. In a sequel, we will focus on incompressible flow, as the formalism would need some revisions for compressible flow.

The problem of finding $D_\epsilon(Re)$ and $1/Re({\cal E})$ are clearly related. In fact, obvious inequalities connect the two problems.

\subsection{Exact bounds}

There are number of exact bounds that link $D_\epsilon$, the dimensionless energy injection, and ${\cal{E}}$, the efficiency. The Cauchy-Schwartz inequality allows us to write
\begin{equation}
D_\epsilon\le \frac{1}{\cal{E}},
\label{Boun1}
\end{equation}
wherein the equality is achieved when $\bm{f}$ is parallel to $\bm{u}$. This mathematical inequality is valid independently of any dynamical considerations. If we further focus on the stationary state, we can use energy balance on the Navier-Stokes equations to infer that $\epsilon$ is equal to the viscous energy dissipation $\epsilon_\nu=\nu\langle \|\nabla\times u\|^2\rangle$, which is positive. Therefore, $\epsilon=\epsilon_\nu>0$ so the absolute value is unnecessary in this specific case.\

The properties of the INSE enable us to derive the following two other inequalities valid in a stationary state (see~\cite{DF02}):
\begin{eqnarray}
\epsilon&\le&c_1\nu \frac{E}{L_f^2}+c_2 \frac{E^{3/2}}{L_f},\\
{\cal E}&\ge&c_3+c_4 Gr^{-1} \left( 1-\sqrt{1+2 c_3 Gr/c_4}\right),
\end{eqnarray}
where $c_i$'s are the non-universal constants that depend on the properties of the external forcing mechanism. Note that the paper~\cite{DF02} claims that the results hold true for the \textit{energy dissipation}, but the proof is really given for the energy injection $\epsilon$, which is equal to the energy dissipation only for non-dissipative stationary weak solutions. Also, the second inequality comes from the Eq.~(50) of ~\cite{DF02}, before the substitution for the Taylor micro-scale is done. We remark that the limit $Gr \to \infty$ implies that the efficiency is bounded from below, ${\cal{E}} \geq c_3$. \

The above observations show that the efficiency cannot vanish in the limit $Gr \to \infty$. If it can diverge in the same limit, the upper bound Eq. (\ref{Boun1}) shows that it would correspond to a solution in which the dimensionless energy injection tends to $0$, while at the same time being able to store an infinite energy at finite forcing, a somewhat physically puzzling result. If it were to tend to a finite value in the limit $Gr \to \infty$, it would correspond to a new interesting law of turbulence. Indeed, Eq.~\eqref{Boun1} indicates that we can have an efficiency saturating in the inviscid limit, while at the same time an energy injection tending to $0$. This shows that the saturation of efficiency and of energy injection/dissipation are different problems.\

Below we discuss several examples that clearly illustrate these different points, on different types of turbulence. We summarise various dimensionless quantities of interest in Table~\ref {tab:def}.

\begin{table}
\begin{center}
\def~{\hphantom{0}}
\begin{tabular}
{ccc}%
 Name& Symbol &Formula\\[3pt]
Reynolds number&Re &$L_f\sqrt{2E}/\nu$ \\
Dimensionless energy injection/dissipation & $D_\epsilon$ &$ \epsilon L_f/E^{3/2}$\\
Efficiency &${\cal E}$ &$EL_f/F_0$\\
\end{tabular}
\caption{Various dimensionless quantities used in this paper as a function of the fluid kinetic energy $E$ and viscosity $\nu$ and forcing intensity $F_0$ and scale $L_f$. }
\label{tab:def}
 \end{center}
\end{table}

\section{Empirical tests of the efficiency saturation}

\subsection{From numerical data}

Numerical data are obtained from numerical simulations of Eq. (\ref{NS}) for two types of viscosities: (i) for time-independent (fixed) viscosity. This corresponds to the classical Navier-Stokes (NS) situation, in which the total kinetic energy $E$ fluctuates dynamically in time. (ii) for time-independent total kinetic energy $E$, which is achieved by monitoring in time 
the viscosity it according to:
\begin{equation}
\nu = \frac{\langle u\cdot f\rangle}{\langle\|\nabla\times u\|^2\rangle}.
\end{equation}
This case is referred to as the \textbf{Reversible Navier-Stokes} (RNS), because it can be shown that such prescription restores the time-reversibility of Eq. (\ref{NS}), which is broken by the addition of a viscosity. The equivalence of the two models on global quantities was originally conjectured by \cite{gallavotti}, and checked in~\cite{shukla2019phase,Guillaume23}, so the two types of data can be used on the same footing. Numerically, simulations of RNSE are more challenging, so that available resolutions are lower. However, the efficiency saturation is easier to control via RNS than NS: instead of varying Reynolds number to report the corresponding variations of energy, it is now possible to impose the energy and obtain the corresponding Reynolds number. It therefore allows for cleaner saturation laws at more modest Reynolds number.

\subsubsection{NS projected on a logarithmic lattice}

We first illustrate the above idea using a simplified model of fluid turbulence obtained by projecting the NS equations onto a Fourier lattice with exponentially spaced modes. The Fourier modes are given by $k_n = k_0 \lambda^n$, where $\lambda$ is the log-lattice parameter (chosen to be $2$ or $\phi = \frac{1+\sqrt5}2$ in our simulations). This projection restricts the number of possible interactions between modes, but obeys the symmetries and the global conservation laws associated with the fluid flow. For a detailed theoretical and numerical discussions we refer to~\cite{campolina2019fluid, Mailybaev18, campolina2021fluid, quentin24}.

\begin{figure}
 \includegraphics[width=0.33\textwidth]{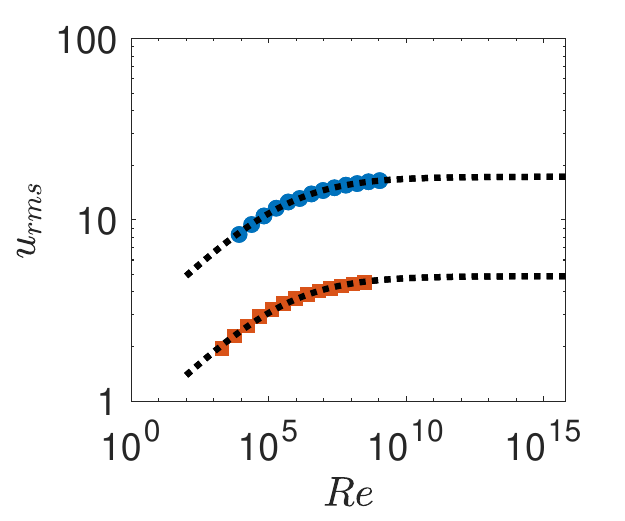}
 \put(-130,125){\bf(a)}
 \includegraphics[width=.33\textwidth]{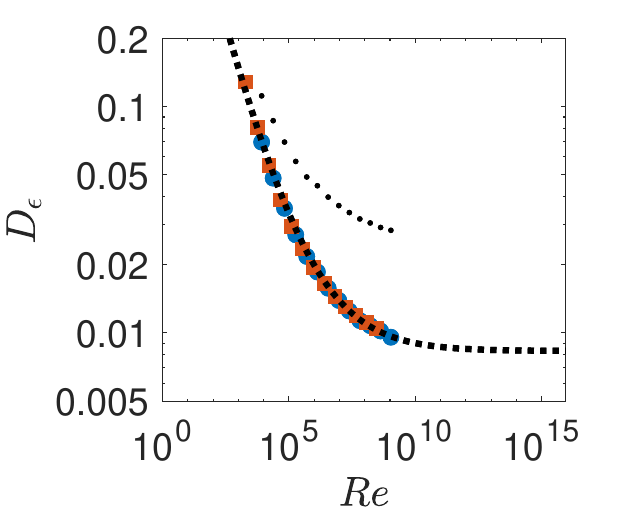}
 \put(-130,125){\bf(b)}
 \put(-75,65){\includegraphics[width=.15\textwidth]{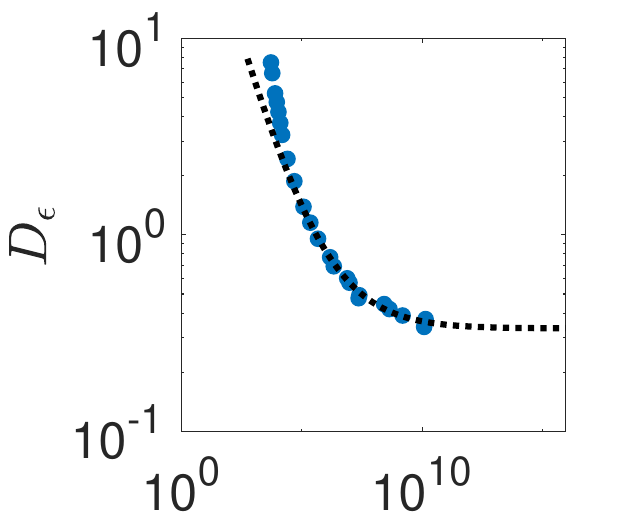}}
 \includegraphics[width=0.33\textwidth]{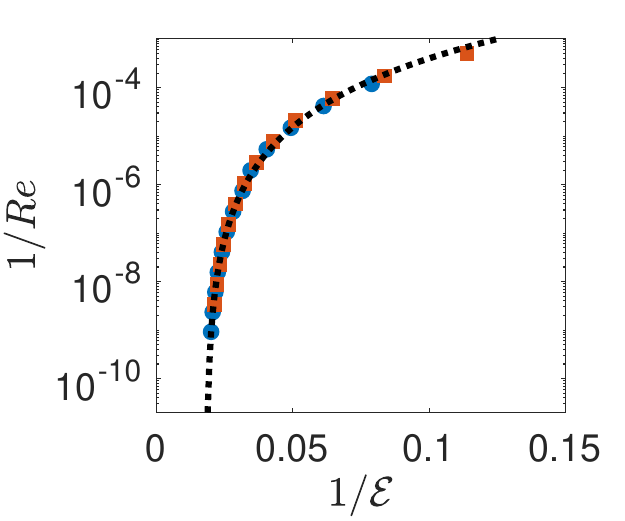} 
 \put(-130,125){\bf(c)}
 \put(-75,25){\includegraphics[width=.15\textwidth]{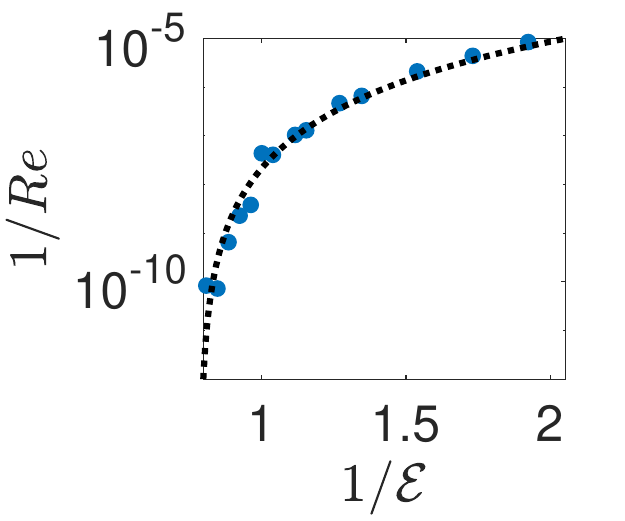}}
 \captionsetup{width = \textwidth, justification=justified}
 \caption{Scaling laws of turbulence on LogLattice: (a) RMS velocity as a function of Reynolds. The data are the symbols, the black dashed line is the fit, assuming a power law for the efficiency, with the corresponding $R^2 = 0.9987$. (b) Dimensionless dissipation as a function of Reynolds. The data are the symbols. The black dotted line is the theoretical bound $\sqrt{2}/{\cal E}$, and the dashed line is the fit $\epsilon_*/{\cal E}^{3/2}$, with the corresponding $R^2 = 0.9960$. (c) Viscosity (inverse Reynolds) versus inefficiency (inverse efficiency). A second order phase transition is observed where viscosity vanishes at finite inefficiency \cite{Guillaume23}. colour codes the forcing type: orange: constant power; blue: constant forcing amplitude. The black dashed line is a fit, assuming a power law for the efficiency, with $R^2=0.9576$. These LogLattice simulations were done with $\lambda=2$ and the insets with $\lambda=\phi$. }
 \label{fig:ToutsurLL}
\end{figure}

In Fig.~\ref{fig:ToutsurLL} we show our results obtained from the log-lattice simulations of Eq.~\eqref{NS} in 3D with a forcing $\bm{f}$ of amplitude $F_0$ around a wave number $k_f=2\pi/L_f$ for different values of viscosity. In these simulations, we use a dyadic spacing $\lambda=2$ and a golden number spacing $\lambda=\phi=1.618$ with a maximal wave number $\kmax$ up to $3 \times 10^6$. We accumulate various statistics of Fourier modes once a statistically stationary state has been achieved. In particular, we define the kinetic energy $E$ and the energy injection as
\begin{equation}
E=\frac{1}{2}\sum_k \overline{u(k,t)u(-k,t)} \quad \text{and} \quad \epsilon=\sum_k \overline{u(k,t)f(-k,t)+u(-k,t)f(k,t)}, \nonumber\\
\label{definitionsutilesE}
\end{equation}
respectively, where the overline denotes a time average. Note that in a stationary state, $\epsilon$ also represents the mean energy dissipation, and $u_{\rm rms}=\sqrt{2E}$ represents the root-mean-squared velocity, since the mean flow is zero. We use these quantities to define the Reynolds number $Re=L_f\sqrt{2E}/\nu$, the dimensionless energy injection/dissipation $D_{\epsilon}=\epsilon L_f / E^{3/2}$ and the efficiency ${\cal{E}}=E L_f / F_0$, also see~Table \ref{tab:def}. 

Figure~\ref{fig:ToutsurLL} (a) shows that despite attaining fairly high $Re$ in our log-lattice simulations, the root mean square (RMS) velocity is yet to reach a saturation, a believed hallmark of fully developed turbulence. Similarly, the dimensionless energy dissipation $D_{\epsilon}$ continues to decrease as $Re$ is increased, however, near the largest $Re$ the decay rate has decreased, but $D_{\epsilon}$ is still far from saturating. Note that the saturation of the energy dissipation to a constant value is a hallmark of the dissipation anomaly. We also observe that for the $Re$ considered here, the bound $D_\epsilon\le \sqrt{2}/{\cal E}$ given by Eq.~\eqref{Boun1}, is always satisfied. In Fig.~\ref{fig:ToutsurLL} (c) we show the behaviour of $1/Re$ vs $1/{\cal{E}}$ for the dyadic lattice. The saturation curve can be very well fitted by an implicit power law as
\begin{equation}
\frac{1}{Re}=A\left({\cal E}^{-1}-{\cal E_*}^{-1}\right)^\gamma,
\label{criticalLaw}
\end{equation}
with $A=10^{-4.86}$, ${\cal{E}_*}=55$ and $\gamma=3.45$. The efficiency saturation in turn provides a dissipation anomaly, following the law with leading order asymptotic expansion (see Fig.~\ref{fig:ToutsurLL} (b)):
\begin{equation}
D_\epsilon = \frac{\epsilon_*}{{\cal E}^{3/2}} = \frac{\epsilon_*}{{\cal E}_*^{3/2}}\left(1+\frac32\frac{\cal E_*}{(ARe)^{1/\gamma}}\right)+o(Re^{-1/\gamma}).
\label{asymptoticLaw}
\end{equation}

Power laws of such type are found in phase transition or in bifurcation, and it is not accidental that we find it here. It was indeed shown in~\cite{Guillaume23} that solutions of the RNS equations projected on log-lattices display a bifurcation at the value ${\cal{E}_*}$: above ${\cal{E}_*}$, only self-similar blow up solutions of Euler equations are observed. Below ${\cal{E}_*}$, dissipative irreversible solutions of the Navier-Stokes equations are observed. Such bifurcation can be observed in the behaviour or the enstrophy, that diverges above ${\cal{E}_*}$, while staying finite below above ${\cal{E}_*}$. Considering the equivalence of NS and RNS, we thus conclude that the power- law we find in our NS case is just the counterpart of the power-law observed in RNS. Note however that in the NS case, we do not have a phase transition, since we observe only one "phase", the dissipative irreversible solutions of Navier-Stokes. The only caveat is that the warm phase where thermalisation occurs is also accessible for NS by using an under-resolved Galerkin truncated DNS.\

The case $\lambda=\phi$ can also be well described by a power law. Specifically, we obtain a good fit for the efficiency using the following set of coefficients $A=10^{-0.85}$, ${\cal{ E}_*}=1.27$ and $\gamma=3.45$, see the inset of Fig.~\ref{fig:ToutsurLL} (c). This may only be two cases, but the different forcing and lattice spacing suggest that the scaling exponents are universal. 

Now, once we obtain the fit, we can invert Eq.~\eqref{criticalLaw} to get $u_{rms}$ as a function of $Re$. We show this in Fig.~\ref{fig:ToutsurLL} (a) using black-dashed lines, the extrapolation to large Reynolds numbers indicates that the saturation is not observed until $Re\sim 10^{12}$, which is considerably beyond our computational capacity. Similarly, we can use this law to fit the variation of the energy dissipation with $Re$ in Fig.~\ref{fig:ToutsurLL} (b) and extrapolate predictions for small viscosities. We find that a good fit to the data is given by $D_\epsilon=\epsilon_*/{\cal E}^{3/2}$, where ${\cal E}(Re)$ is obtained by inverting Eq.~\eqref{criticalLaw} with constant $\epsilon_*=53$ (resp. $\epsilon_*=8$) for $\lambda=2$ (resp. $\lambda=\phi$). The extrapolated black-dashed curve suggests that the dimensionless energy dissipation becomes constant in the limit of vanishing viscosity, which is consistent with the existence of dissipation anomaly (the so-called zeroth law of turbulence). 

\subsubsection{Direct Numerical simulation of RNS}

We now repeat the analysis for the homogeneous isotropic turbulence obtained using the DNS's of the time-reversible Navier-Stokes equations (RNS) over a triply period domain of size $(2\pi)^3$. For more details we refer to~\cite{shukla2019phase}. In RNS, the imposition of a global constraint of the conservation of a quadratic quantity, for example, the total energy, allows for the viscosity $\nu(t)$ to fluctuate in time, however, its statistical properties are equivalent to those obtained using the standard INSE Eq.~\eqref{NS}. We use two different codes: $S1$ a classical pseudo-spectral code with $64^3$ and $128^3$ collocation points, $S2$ a pseudo-spectral code with Taylor-Green symmetries that allows us to perform DNS's with $512^3$ and $1024^3$ collocation points. Note that the Reynolds numbers reached here are considerably smaller than those of log-lattice simulations. 

\begin{figure}
 \includegraphics[width=0.33\textwidth]{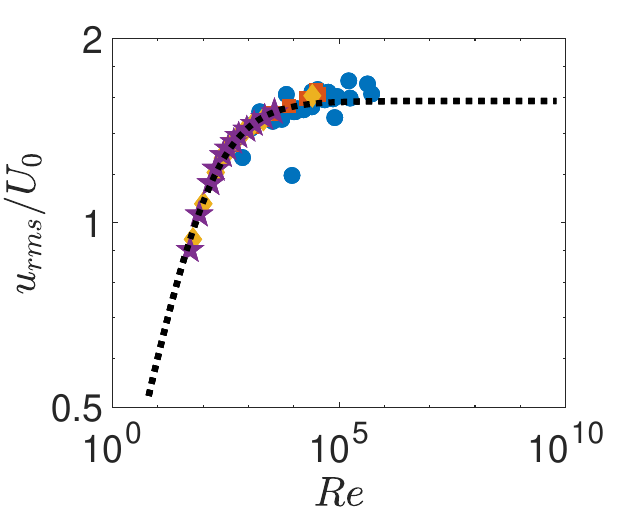}
 \put(-130,125){\bf(a)}
 \includegraphics[width=.33\textwidth]{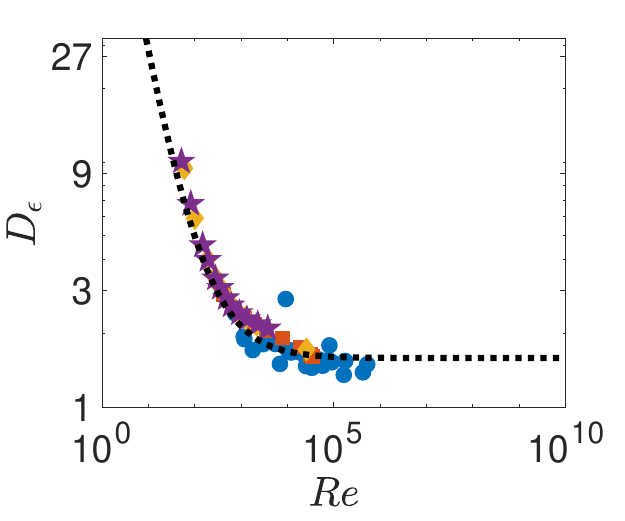}
 \put(-130,125){\bf(b)}
 \includegraphics[width=0.33\textwidth]{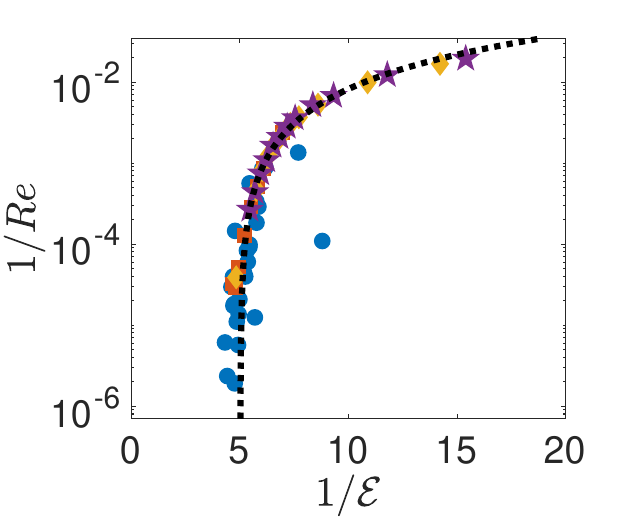} 
 \put(-130,125){\bf(c)}
 \captionsetup{width = \textwidth, justification=justified}
 \caption{Scaling laws of turbulence in numerical homogeneous isotropic turbulence: (a) RMS velocity as a function of Reynolds. To account for different forcing type, we have divided the RMS velocity by $U_0=\sqrt{F_0 L_f}$. The data are the symbols, the black dashed line is the fit, assuming a power law for the efficiency, with $R^2 = 0.9844$. (b) Dimensionless dissipation as a function of Reynolds. The data are the symbols. The dashed line is the fit $2.5/{\cal E}^{3/2}$, with $R^2= 0.8733$ (c) Viscosity (inverse Reynolds) versus inefficiency (inverse efficiency). Red and yellow symbols: S1 at resolution $64^3$ and $128^3$, magenta and blue symbols: S2 at resolution $512^3$ and $1024^3$. The black dashed line is a fit, assuming a power law for the efficiency, with $R^2=0.9706$. $Re$ is not explored as far as in LogLattice, but the phase transition/bifurcation begins at smaller $Re$, with a larger inefficiency compared to LogLattice.}
\label{fig:ToutsurDNS}
\end{figure}

Figure~\ref{fig:ToutsurDNS} (a) and (b) show that despite the comparatively smaller Reynolds numbers both $u_{rms}/U_0$ and the dimensionless dissipation, respectively, begin to saturate for $Re \simeq 10^4$. Similarly $1/Re$ starts to show saturation starting ${\cal{E}}_* \sim 0.2$, see Fig.~\ref{fig:ToutsurDNS} (c). A fit of the form Eq.~\eqref{criticalLaw} is obeyed with the exponent $\gamma=1.4$ and the prefactor $A=10^{-0.2}$. Inversion of this power law allows us to obtain a parameter-free fit of the kinetic energy in Fig.~\ref{fig:ToutsurDNS} (a) and a one parameter fit of the energy dissipation $2.5{\cal{E}}^{-3/2}$. In summary, the DNS's results are very similar to the case of the flows projected on log-lattices, but belong to a different "universality class" (laws have the same shape but different constants). Given the equivalence between RNS and NS, we can trace this difference to number of allowed triadic interactions in Fourier space.

\subsubsection{Comment on the dissipation scaling}
Through the efficiency lens, we are able to connect energy dissipation and to the Reynolds number by inverting the power law, resulting in Eqs. (\ref{asymptoticLaw}).
It is instructive to compare the asymptotic behaviour of the dissipation anomaly with other simple models. A qualitatively similar formula can be obtained by a mean field approach:
\begin{equation}
D_\epsilon = D_{\epsilon,\infty}\left(\sqrt{1+\frac{3b^3}8\frac1{Re^2}}+\sqrt{\frac{3b^3}8}\frac1{Re}\right) = D_{\epsilon,\infty}\left(1+\sqrt{\frac{3b^3}8}\frac1{Re}\right)+o(Re^{-1}),
\end{equation}
where $b$ is the Kolmogorov's constant for the two-point structure constant. A full derivation can be found in~\cite{Lohse_1994}. This theoretical approach has the benefit of predicting that $\gamma=1$. This is not specific to the chosen closure model of the article akin to a turbulent viscosity:
\begin{equation}
\nu_t = \frac{D^2}{b^3\epsilon}
\end{equation}
where $D$ is the two-point structure function. Any closure at the level of the two-point correlation function will necessarily give $\gamma=1$. This follows from the K\'arm\'an-Howarth equation itself. In the inviscid case, the third order structure function expressed in terms of $D,\epsilon$ but independent of $\nu$ (given by the choice of closure), balances the dissipation term, with a finite dissipation rate solution. The viscous term is proportional to $\nu$, and will therefore generically give a $O(\nu)=O(Re^{-1})$ correction to the asymptotic dissipation rate. In statistical mechanics, the analogue would be the Landau theory whose structural limitations only allow it to predict mean field critical exponents. To reproduce the numerical data, where $\gamma=3.4$ for LogLattice or $\gamma=1.4$ for DNS, the full hierarchy of structure functions seem to be necessary, requiring at least some analogue of the renormalisation group to recover the anomalous scaling.

\subsection{From experimental data}
So far, all the results were obtained in well controlled situations, where all the quantities are known exactly. We now turn to the case of experimental flows, to show how the efficiency can be computed from classical global diagnostics of the flow and possible incomplete data.

\subsubsection{Von K\'arm\'an flows}

\begin{table}
\begin{center}
\def~{\hphantom{0}}
\begin{tabular}
{ccc}%
 Name & Symbol & Description / Formula\\[3pt]
Radius &$R_d,R$ & impeller, tank radius \\
Height & $H$ & distance between impellers ($H/R = 1.8$) \\
Frequency & $F$ & common rotation frequency of impellers \\
Torque & $\Gamma$ & measured response of the flow\\
Dimensionless torque & $K_p$ & $ \frac{\overline{\Gamma}}{ \rho \pi R^5 (2\pi F)^2}$ \\
Dimensionless Energy & $E$ & $\frac{\langle \bm{u}\cdot \bm{u}\rangle^{1/2}}{2 (2\pi R F)^2}$
\end{tabular}
\caption{Parameters of the von K\'arm\'an experiment.}
\label{tab:vk}
 \end{center}
\end{table}

First, we consider the Von K\'arm\'an flow generated between two counter or co-rotating impellers in a cylindrical tank. The experimental setup has impellers of radius $R_d$ in a tank of radius $R$ filled with a fluid of density $\rho$ and viscosity $\nu$. The distance between the two impellers is $H=1.8 R$. The fluid is forced and maintained in a turbulent state at a high Reynolds number by means of the two independently rotating impellers at the same constant frequency $F$, located at the top and the bottom of the cylindrical tank. Flows with different Reynolds numbers are achieved by using different fluids: pure glycerol, mixture of glycerol and water, and water at $T=20$ $^{\circ}$C. In all the cases, the temperature is kept constant through a thermal regulation, to avoid drifts of viscosity during the operation. The mean kinetic energy of the flow can be computed on a meridian plane, including the rotation plane, with the help of stereoscopic particle image velocimetry (SPIV) technique, which allows for the reconstruction of the three components of the velocity within this plane. For further technical details about these measurements, we refer to~\cite{thesemonchaux,thesedebue,cortet,diribarne}. 

The axisymmetry of the flow ensures that the planar estimate is equal to the volume average of the kinetic energy. Moreover, inserting two torque meters, before and below the top and bottom impellers, respectively, allows for the measurement of the torque $\Gamma(t)$ applied by the turbulent flow on the impeller; this provides an estimate of the amplitude of the forcing per unit mass required to maintain the flow in a stationary state. The energy dissipation rate is measured using the mean injection rate, obtained by computing the mean work done by the torque $W=\overline{\Gamma(t) F}$, where the overline denotes a time average. The choice of the unit of length $L=R$ and the unit of velocity $U=2\pi RF$, allows us to combine the parameters summarised in Table~\ref {tab:vk} into the mean torque $K_p$, the kinetic energy $E$ and the global Reynolds number $Re_F$ in dimensionless form in as follows
\begin{equation}
K_p = \frac{\overline{\Gamma}}{ \rho \pi R^5 (2\pi F)^2}, \quad E = \frac{\langle \bm{u}\cdot \bm{u}\rangle^{1/2}}{2 (2\pi R F)^2}, \quad \text{and} \quad Re_F = \frac{2\pi R^2 F}{\nu},
 \label{defiVK1}
\end{equation}
respectively, where $\langle\cdot\rangle$ denotes an average over the meridian plane. In that experimental configuration, the mean torque provides two points of information: first, it directly provides the mean force applied to the flow by the propellers, providing an estimate of $F_0$. Second, it can also provide the energy injection/dissipation $\epsilon$ in experimental conditions where the velocities of the upper and lower impellers are maintained constant in time. Indeed, in such a case, the mean power injected at the top and the bottom is computed as the time-average of the torque times the frequency of the impeller. The latter being constant, the power is just the product of the mean torque times the (constant) frequency. We are thus in a situation where the equality in Eq. (\ref{Boun1}) is realised, so that the two quantities tend to zero or saturate together. 

 \begin{figure}
 \centering
 \includegraphics[width=.33\textwidth]{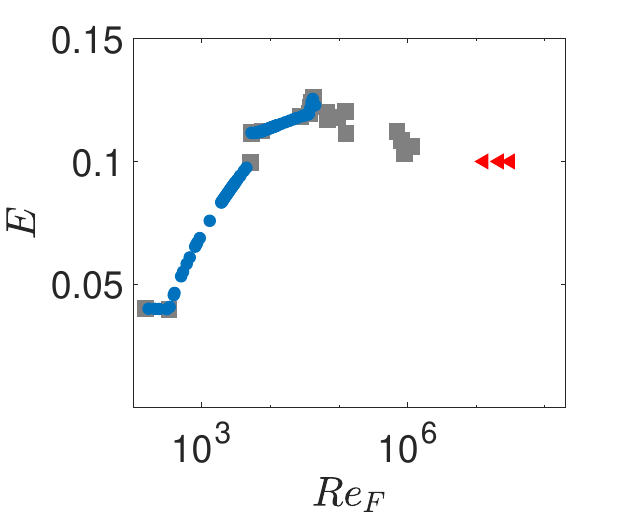}
 \put(-130,125){\bf(a)}
 \includegraphics[width=.33\textwidth]{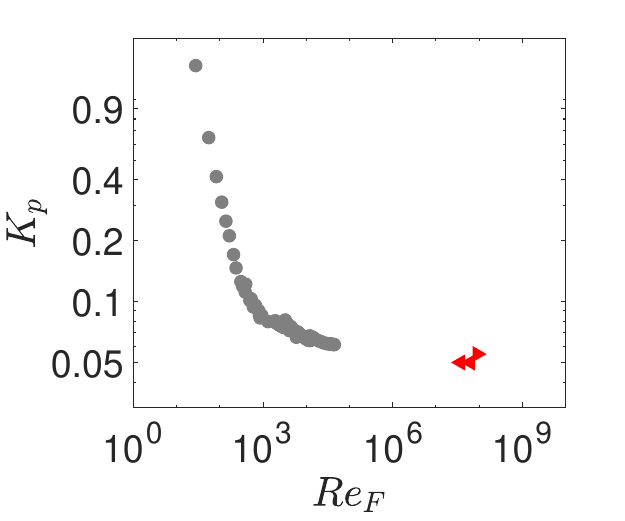}
 \put(-130,125){\bf(b)}
 \includegraphics[width=.33\textwidth]{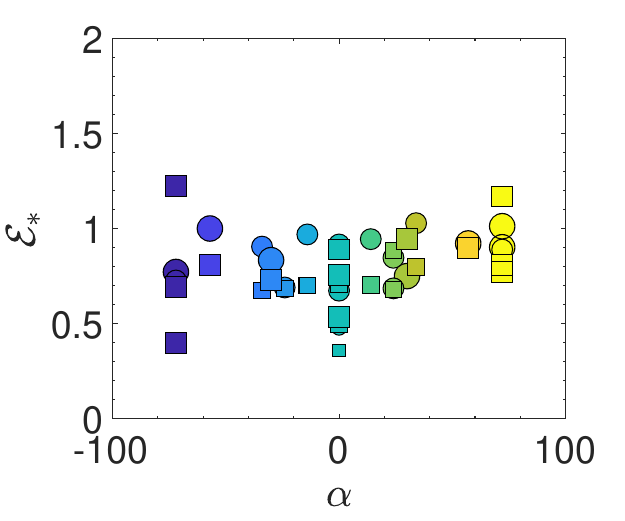}
 \put(-130,125){\bf(c)} 
\captionsetup{width = \textwidth, justification=justified}
\caption{Measurements in the Von K\'arm\'an flow with counter-rotating impellers. (a) Kinetic energy $E$ vs global Reynolds number $Re_F$. Grey dos correspond to the real data from PIV, while the blue data interpolates the missing data using a supercritical law $E \sim (Re_F-3500)^{1/2}$~\citep{raveletjfm}. (b) Mean dimensionless torque $K_p$ vs $Re_F$ as measured by \cite{raveletPhD} (grey dots). The red triangles correspond to points measured in the SHREK experiments (superfluid Helium) using 8 bladed impellers with curvature $72^\circ$. (c) Efficiency of impellers of various radius, fitted with blades of curvature angle $\alpha$. }
\label{fig:DispoSurVK}
\end{figure}

\begin{figure}
 \includegraphics[width=.49\textwidth]{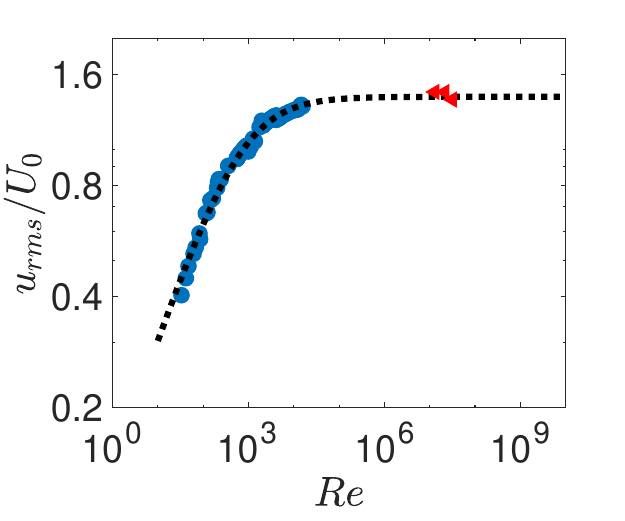}
 \put(-200,180){\bf(a)}
 \includegraphics[width=.49\textwidth]{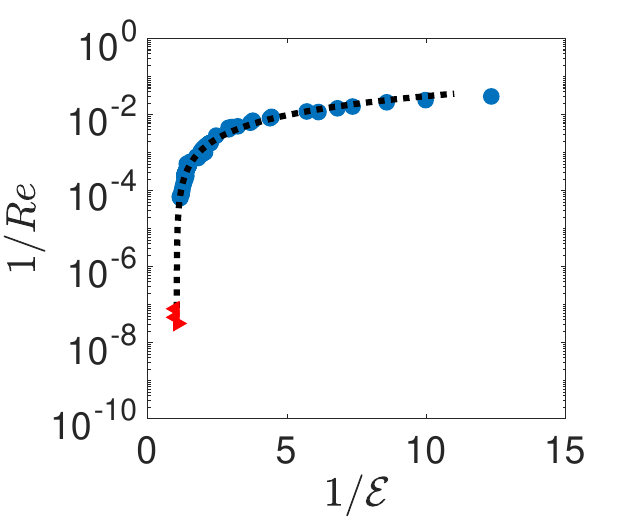} 
 \put(-200,180){\bf(b)} 
 \captionsetup{width = \textwidth, justification=justified}
\caption{Scaling laws of turbulence in experimental Von K\'arm\'an flow with counter-rotating TM60+ impellers: (a) RMS velocity as a function of Reynolds. To account for different forcing amplitude, we have divided the RMS velocity by $U_0=\sqrt{F_0 L_f}$. The data are the blue symbols, the black dashed line is the fit, assuming a power law for the efficiency, with $R^2= 0.9915$. ( (b) Viscosity (inverse Reynolds) versus inefficiency (inverse efficiency). The red triangles correspond to points measured in the SHREK experiments (superfluid Helium) using 8 blades impellers \citep{shreksm}. The black dashed line is the fit, assuming a power law for the efficiency, with $R^2= 0.9730$. }
\label{fig:ToutsurVK}
\end{figure}

We analyse the data from the experimental runs, wherein the counter-rotating impellers are fitted with $16$ curved blades of angle $72^\circ$ in the scooping direction and results in a symmetrical large scale circulation~\citep{raveletPhD,cortet}; such impellers are referred to as TM60+. In Fig.~\ref{fig:DispoSurVK} (a), (b) and (c) we show the plots of $E$, $K_p$ and ${\cal{E}_*}$, respectively, using the available data. We have sparse data points for the kinetic energy in the Reynolds number range $Re_F \in [10^2, 10^6]$ and a detailed set of measurements of the dimensionless torque $K_p$ for the range $Re_F \in [10^2, 10^5]$. To be able to combine both the measurements, we first interpolate the kinetic energy over the range of Reynolds number for which the detailed measurements of $K_p$ exists. We perform the interpolation using the information that the transition from the laminar to turbulent state is supercritical, with an exponent 2 and critical Reynolds number $Re_*=3500$, see~\cite{raveletjfm}. Following this, we compute all the quantities of interest (in dimensionless units)
\begin{equation}
F_0 = \frac{K_p R^2 L_f }{ R_d^2}, \quad u_{rms} = \sqrt{2E}, \quad \text{and} \quad R_{rms} = u_{rms}L_f Re_F,
\label{defiVK2}
\end{equation}
where $L_f$ is the dimensionless forcing scale. For the rest of this discussion, we take $L_f$ to be unity. We show the results of the analysis similar to the numerical data in Fig.~\ref{fig:ToutsurVK}. We do not observe any clear saturation of the quantities $u_{rms}/U_0$, $D_{\epsilon}=1/{\cal E}$ and $1/Re$. However, if we assume a power law for the efficiency~\eqref{criticalLaw}, we find that a good fit for the parameters $A=10^{-2.7}$, ${\cal{E}}_*=0.97$ and $\gamma=1.4$, which is similar to the homogeneous isotropic turbulence case. Furthermore, we can use this fit to extrapolate the behaviour at higher Reynolds numbers. However, the above discussed experimental runs do not have torque data at further higher Reynolds numbers to check the quality of the extrapolation for the efficiency. Therefore, we use the torque data at very large Reynolds numbers measured in the SHREK facility~\citep{shrekdiribarne, shrekrousset, shreksm}. Note that the SHREK setup has the same geometry as the experimental setup discussed above, except that the impellers are fitted with $8$ blades instead of $16$. We show their data in Fig.~\ref{fig:ToutsurVK} using red filled triangles, which lie on the extrapolated behaviour (black dashed line).

We check the dependence of ${\cal{E}_*}$ with respect to the forcing details. To achieve this, we compute the efficiency at $Re_F \sim 10^6$ based on the measurements of $K_p$ and the kinetic energy as reported in~\cite{raveletPhD}, wherein these quantities were obtained for impellers of various radius and fitted with blades of different curvature angle $\alpha$. We find that the data is somewhat scattered, but on an average lie around the value ${\cal{E}_*} \sim 1$, which shows that this value is independent of this geometry. This is rather remarkable, since the value of $\alpha$ has a profound impact on the flow anisotropy, as measured by the ratio between the poloidal to the toroidal energy \citep{raveletPhD}.

\subsubsection{Pipe flows}

The second experimental flow that we consider is an example of a wall bounded parallel flows. These are characterised by a mean flow that depends only on one space coordinate, as a consequence streamlines are parallel. We choose a circular pipe flow and analyse the data available from the experimental works of~\cite{MKJMS04,SJID02}. The flow is driven by a pressure gradient along the axis of the pipe, $\partial_x P$, and results into a mean velocity $\overline{U}$. In smooth pipes, pressure gradients balance the shear stress, so that the force applied to the flow can be measured either through the pressure drop along the pipe or through the friction drag $\tau_{w}$. For the given fluid density $\rho$, the viscosity $\nu$ and the pipe diameter $D$, we can build two dimensionless parameters
\begin{equation}
\lambda = 8\frac{-(\partial_x P)D}{ \rho \overline{U}^2} \quad \text{and} \quad Re = \frac{\overline{U} D}{\nu}.
\label{defiWBF}
\end{equation}

\begin{figure}
\centering
	\includegraphics[width=.45\textwidth]{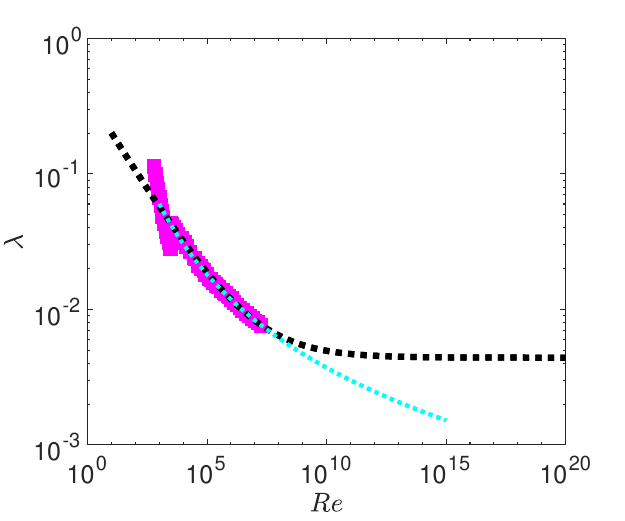} 
 \put(-200,180){\bf(a)} 
 \includegraphics[width=.45\textwidth]{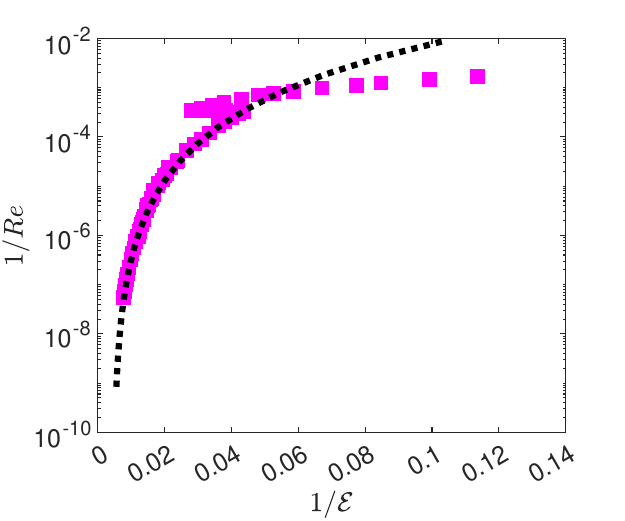} 
 \put(-200,180){\bf(b)} 
 \captionsetup{width = \textwidth, justification=justified}
\caption{(a) Friction factor as a function of $Re$ in pipe flow. The black dotted line is obtained by inverting the power law for efficiency, with $R^2= 0.9936$. The blue dotted line is the Prandtl-type formula with $R^2= 0.9966$. The quality of the fit shows that for this data both laws have similar accuracy. Only experimental data at much higher Reynolds number could discriminate between the two. (b) Check of the power law for the efficiency. The dotted line is a fit with the power law of slope $3.5$ and $R^2=0.9966$.}
\label{fig:ToutsurWF}
\end{figure}

We show the plots of $\lambda$ vs $Re$ and $1/Re$ vs ${\cal{E}}$ in Fig.~\ref{fig:ToutsurWF} (a) and (b), respectively. The force applied to the flow is $F_0=(1/\rho) \partial_x P$. We further assume that the typical forcing wave number is $k_f=1/D$. To compute the efficiency, we need the kinetic energy $E$ stored in the flow. It can be split into two contributions, following $E=\overline{U}^2/2+ u^2_{rms}/2$. The first term is due to the presence of a mean flow velocity and the second term is associated with the RMS velocity $u_{rms}$. However, we have $u^2_{rms} \ll \overline{U}^2$ because of the presence of a large mean flow. As a consequence we can approximate the kinetic energy as $E=\overline{U}^2/2$. Therefore, with this convention, we have $\lambda \sim 1/{\cal{E}}$. In Fig.~\ref{fig:ToutsurWF} (b) we show the behaviour of $1/Re$ vs $\lambda$ (or ${\cal{E}}$), where we observe that it follows the power law~\eqref{criticalLaw} with an exponent $\alpha=3.5$. Clearly, this exponent is larger than the one that we obtained for the homogeneous isotropic turbulence, which shows shows a sensibility to anisotropy and inhomogeneity.

When we invert the efficiency power law, we observe a saturation in the behaviour of $\lambda$, see Fig.~\ref{fig:ToutsurWF} (a). This is in stark contrast to the Prandtl formula which predicts a decaying behaviour
\begin{equation}
\lambda=\left[0.8382\;W(0.6287\; Re)\right]^{-2} \sim \frac1{(0.8382\ln Re)^2},
\end{equation}
where $W$ is the Lambert function~\citep{Dubrulle_2024}. The fit displayed in Fig.~\ref{fig:ToutsurWF} (a) shows that effect of saturation will only be felt at $Re\sim 10^{12}$ a value way too large for present experimental facilities.

\subsubsection{Taylor-Couette flow}
Taylor-Couette flow is obtained by the co or counter-rotation of two independent co-axial cylinders, with radii $a$ for the inner cylinder, and $b$ for the outer cylinder. Many different experiments have been built to study such flow \citep{paoletti12}. We use here data from the Twente turbulent TC facility T$^3$C, kindly provided to us by D. Lohse. The experiment is described in great detail in \cite{vangils11b}. It is characterised by cylinders of height $L = 92.7$~cm, and respective radii $a = 20.0$~cm and $b = 27.94$~cm. The maximum inner and outer angular velocities are $\Omega_1/2\pi = 20$~Hz and $\Omega_2/2\pi = \pm 10$~Hz, respectively. The torque is measured over the middle part of the inner cylinder of height $L_{mid} = 53.6$~cm to minimise the influence of the end-plates (similar to \cite{lathrop92a,lathrop92b}). This allows for the definition of the dimensionless torque $G = T/\rho \nu^2 L_{mid}$, where $T$ is the torque, $\rho$ the fluid density and $\nu$ the kinematic viscosity.\

The Taylor-Couette flow is characterised by 
 three essential parameters \citep{dubrulle05}, namely:
 \begin{itemize}
 \item its Reynolds number $Re=
\frac{2}{1+\eta}\vert \eta Re_2 - Re_1\vert$;
\item its Rotation number $R_\Omega=
(1-\eta)\frac{Re_1 + Re_2}{\eta Re_2 - Re_1}$;
\item its curvature number $R_{\cal C}=\frac{1-\eta}{\eta^{1/2}}$;
\end{itemize}
where $Re_1=a\Omega_1d/\nu$ and $Re_2=b\Omega_2d/\nu$ are the Reynolds number of the inner and outer cylinder.
The above control parameters have been introduced so that their definition apply to rotating shear flows in general and not only to the Taylor-Couette geometry. It is very easy in this formulation to relate the Taylor-Couette flow to shearing sheet (the plane Couette flow with rotation), by simply considering the limit $R_{\cal C}\rightarrow 0$. \

Torque measurements provide the force intensity applying to the turbulent flow. Moreover, like in the von K\'arm\'an case, its also provides the power through the time average of the product of torque times the frequency. Since frequencies of the inner and outer cylinders are time-independent, we are in a situation where efficiency and energy injection follow the equality in Eq. (\ref{Boun1}).\

\begin{figure}
\centering
	\includegraphics[width=.45\textwidth]{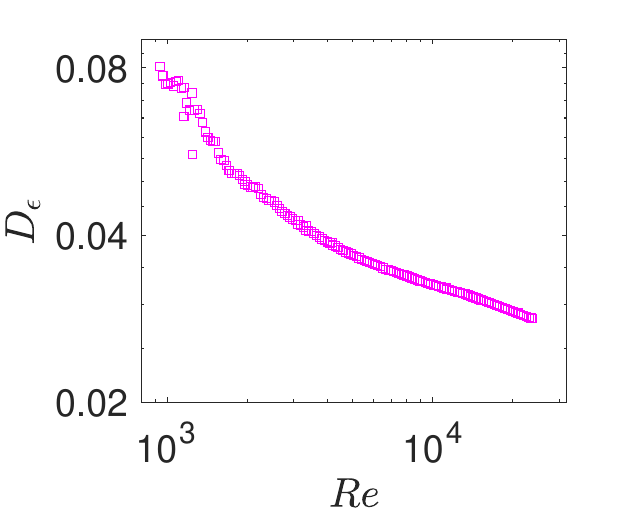} 
 \put(-200,180){\bf(a)} 
 \includegraphics[width=.45\textwidth]{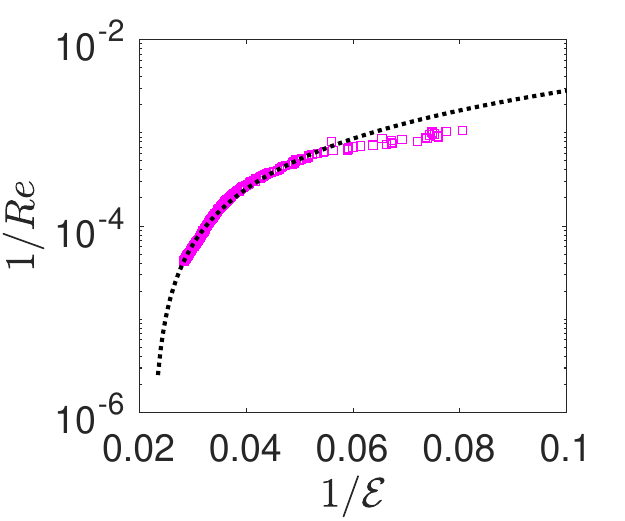} 
 \put(-200,180){\bf(b)} 
 \captionsetup{width = \textwidth, justification=justified}
\caption{Reduced efficiency and friction factor in Taylor-Couette flow, obtained in a situation where only the inner cylinder is rotating. (a) Friction factor as a function of $Re$ in Taylor-Couette flow. (b) Check of the power law for the efficiency. The dotted line is a power law with exponent $1.634$. The fit was performed using only fully turbulent cases, i.e. $Re>2\cdot 10^3$, giving $R^2=0.9892$. Data from the Twente turbulent TC facility T$^3$ \citep{vangils11b}. Data courtesy D. Lohse.}
\label{fig:ToutsurTC}
\end{figure}

Applying the definition (\ref{Re&E}), we thus obtain that in the case of the Taylor-Couette flow, the efficiency and the dimensionless dissipation are given by ${\cal E}=\frac{1}{D_\epsilon}=\frac{Re^2}{G}$. Saturation of the efficiency or of the dimensionless dissipation correspond to a situation where $G\propto Re^2$. This is called the ultimate regime, in which transport properties do not depend anymore on the molecular viscosity. Detailed analysis of $G$ as a function of $Re$ and $R_\Omega$ have been performed in \cite{dubrulle05,paoletti12}. It was found that for all existing experimental data, it was possible to factorise the torque as $G=f(R_\Omega)G_i$, where $f$ is a function that only depends on $R_\omega$ and $G_i$ is the torque computed in a situation where the outer cylinder is at rest. Using this property, we can also factorise the efficiency and the dimensionless energy injection (friction factor) as:
\begin{equation}
{\cal E}=\frac{1}{D_\epsilon}=\frac{1}{f(R_\Omega)}\frac{Re^2}{G_i}.
\label{effiTC}
\end{equation}
This shows that it is sufficient to consider only the behaviour of $Re^2/G_i$ as a function of $Re$. This behaviour is reported in Figure \ref{fig:ToutsurTC}, for a case where only the inner cylinder is rotating. One sees that the friction factor does not saturate, but steadily decays with increasing $Re$, in the range of $Re$ considered. Even though the range of Reynolds number is not very large, we have tried to see whether the reduced efficiency could be fitted by a power law Eqs. (\ref{criticalLaw}) with $A=10^{-0.73466}$, ${\cal{E}_*}=44$ and $\gamma=1.634$. The result is shown in Figure \ref{fig:ToutsurTC}-b. The fit is clearly not as good as for previously considered cases, and it may well be that the efficiency never saturates in this case. Nonetheless, more data would be needed to conclude.

\subsubsection{Rayleigh-B\'enard convection}
\begin{figure}
 \includegraphics[width=0.33\textwidth]{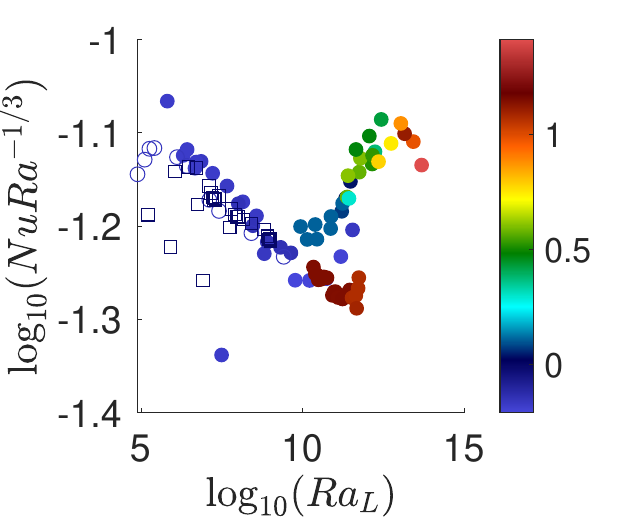}
 \put(-130,125){\bf(a)}
 \includegraphics[width=.33\textwidth]{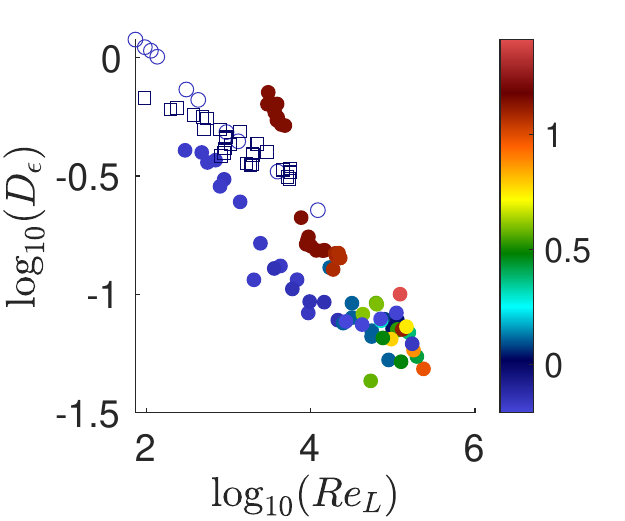}
 \put(-130,125){\bf(b)}
 \includegraphics[width=0.33\textwidth]{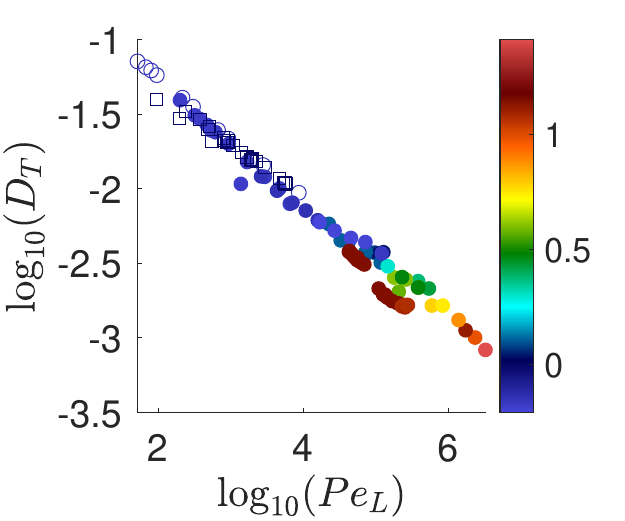} 
 \put(-130,125){\bf(c)}
 \captionsetup{width = \textwidth, justification=justified}
 \caption{Scaling laws of turbulence in experimental and numerical Rayleigh-B\'enard convection. To account for different aspect ratio, we have used everywhere the effective length scale $H_{eff}=H(\Gamma^2/(1/49+\Gamma^2)^{1/2}$ \citep{ABH++22}. Filled data: experiments by \cite{chavanne} and \cite{MBSC21}. Open square: 
 numerical data by \cite{SBZVL18}; open circle: numerical data by \cite{SS17}. The data are coloured according to $\log_{10}(Pr)$. (a) Rescaled dimensionless heat transfer as a function of Rayleigh. (b) Dimensionless kinetic energy dissipation as a function of Reynolds. (c) Dimensionless thermal energy dissipation as a function of P\'eclet. Data courtesy O. Shishkina.}
\label{fig:ToutsurRBC}
\end{figure}

The Rayleigh-B\'enard (RB) setting corresponds to a situation where the fluid is heated from below, and enclosed between two plates, separated by a distance $H$. The temperature difference between the two plates is $\Delta T$. This classical setting has been the topic of many numerical and experimental studies \citep{roche2020ultimate,LS2024}. This is a paradigmatic example of a flow forced by buoyancy. The RB flow is characterised by three dimensionless control parameters
namely:
 \begin{itemize}
\item its Rayleigh number $Ra=\frac{g\beta \Delta T H^3}{\kappa\nu}$;
\item its Prandtl number $Pr=\frac{\nu}{\kappa}$;
\item its aspect ratio $\Gamma=\frac{H}{L}$;
\end{itemize}
and two dimensionless response parameters:
 \begin{itemize}
\item its Reynolds number $Re= \frac{UH}{\nu}$; 
\item its Nusselt number (dimensionless heat flux $Nu=J H/\kappa \Delta T $).
\end{itemize}
where $J$ is the heat flux, $U=\sqrt{2E}$ is the typical velocity, $g$ the gravity acceleration, $\beta$ the coefficient of thermal expansion, $\nu$ the kinematic viscosity, $\kappa$ the thermal diffusivity and $L$ a typical horizontal scale. In the case of no-slip boundary conditions, the variation of $Nu$ and $Re$ can been approximated with effective scaling laws,
$Nu\sim Pr^\alpha Ra^{\gamma}$, for which the prefactor and effective exponent depends on the range in $Ra-Pr-\Gamma$ parameter space. All experiments and numerical simulations reported so far show 
$\gamma<1/2$ \citep{LS2024}. Through a recent theoretical and experimental effort, the aspect ratio dependence of the heat transfer in cylindrical volume of various aspect ratio has been elucidated in \cite{ABH++22}, and shown to be amenable to the introduction of a new effective height $H_{eff}=H(\Gamma^2/(1/49+\Gamma^2)^{1/2}$. An example of data collapse of different aspect ratios and $Pr$ is shown in Figure (\ref{fig:ToutsurRBC})-a for numerical \citep{SBZVL18,SS17} or experimental data \citep{chavanne,MBSC21} at various $\Gamma$ and $Pr$. They show a mild decay of $Nu/Ra^{1/3}$, followed by a steeper rise at larger Rayleigh numbers, evidencing the transition between $\gamma<\sim 1/3$ to $\gamma>1/3$ as the Rayleigh number increases. These data have been selected because they offer measurements of both $Nu$ and $Re$ as a function of $\Gamma$, $Pr$ and $Ra$, a feature that will be important in the computation of the efficiency, see below. However, they are typical of all data gathered so far, see \cite{ABH++22} for example of collapse with more data.\

Theories explaining scaling laws in RB flow are generally based on two exact relations for the kinetic and thermal energy dissipation \citep{SS90,LS2024}:
\begin{eqnarray}
\epsilon_k &=\nu^3(Nu-1) Ra Pr^{-2}/H^4, \quad \epsilon_t &=\kappa (\Delta T)^2 Nu/H^2
\label{exactSSBC}
\end{eqnarray}
or, in dimensionless form:
\begin{eqnarray}
D_\epsilon&=\frac{\epsilon_k H}{U^3}=(Nu-1) Ra Pr^{-2} Re^{-3},\quad D_T&=\frac{\epsilon_t H}{(\Delta T)^2 U}=\Nu Re^{-1} Pr^{-1}.
\label{exactSSBCnondim}
\end{eqnarray}
Their behaviour as a function of $Re$ or $Pe=Re Pr$ is shown in Figure (\ref{fig:ToutsurRBC})-b and (\ref{fig:ToutsurRBC})-c. In both cases, they decay steadily, showing no evidence of saturation. Saturation of both $D_\epsilon$ and $D_T$ would correspond to the ultimate regime of convection, in which $Nu\sim RePr\sim (Ra Pr)^{1/2}$.\

\begin{figure}
\centering
	\includegraphics[width=.45\textwidth]{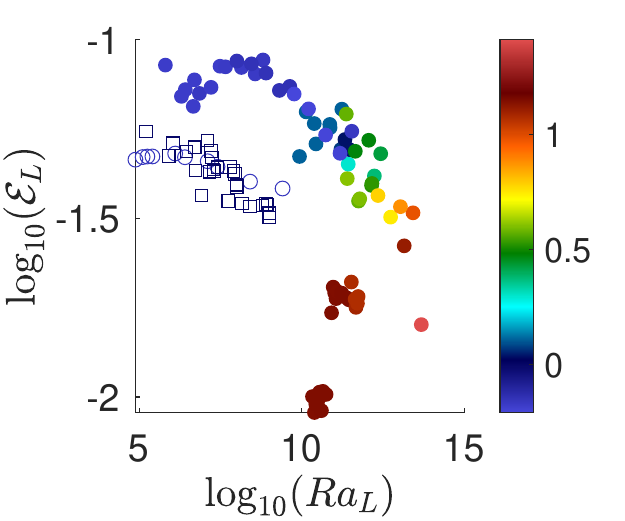} 
 \put(-200,180){\bf(a)} 
 \includegraphics[width=.45\textwidth]{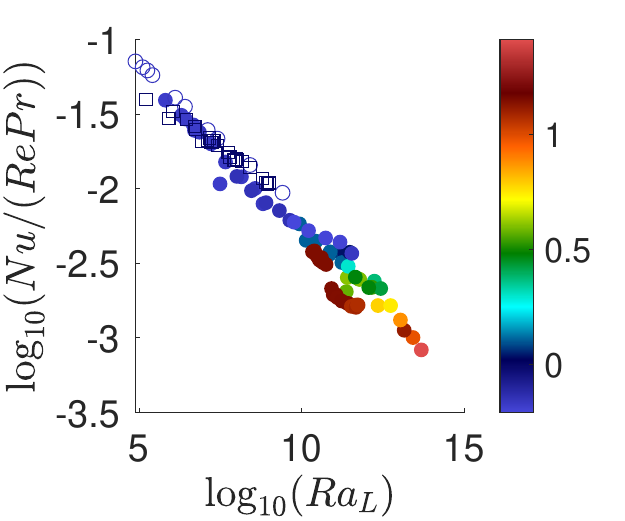} 
 \put(-200,180){\bf(b)} 
 \captionsetup{width = \textwidth, justification=justified}
\caption{Tests of ultimate regime in Rayleigh-B\'enard convection. (a) Efficiency as a function of $Ra$. (b) $Nu/RePr$ as a function of $Ra$. Same symbols and colours than in Figure (\ref{fig:ToutsurRBC}). Data courtesy O. Shishkina.}
\label{fig:EffiRBC}
\end{figure}

The forcing being due to the buoyancy, we have $F_0=g\beta \Delta T$. Therefore, the efficiency can be simply expressed as a function of the Rayleigh and Reynolds number as
\begin{eqnarray}
{\cal E}=Re^2 PrRa^{-1}.
\end{eqnarray}
Injecting this into Eq. (\ref{exactSSBCnondim}), we then get $D_\epsilon=(Nu-1)/(RePr{\cal E})$. Thanks to the exact bound $Nu<Re Pr$ \citep{LS2024}, we see that $D_\epsilon$ is indeed bounded by $1/{\cal E}$, in accordance with Eq. (\ref{Boun1}). This analysis therefore provides an alternative interpretation of the required conditions to get the ultimate, asymptotic regime, as being achieved when both $Nu/(RePr)$ and ${\cal E}$ saturate. The verification of such saturation is provided in Figure (\ref{fig:EffiRBC}). We observe no sign of saturation for either quantity at the Rayleigh numbers investigated.\

To investigate the inviscid limit of $D_\epsilon$ and ${\cal E}$, one can use the extension of the Grossmann and Lohse theory to the ultimate regime \citep{LS2024}. This theory identifies 4 different asymptotic regimes (for $\Ra\to \infty$) depending on the scaling of $Pr$ with $Ra$ namely:
\begin{itemize}
\item Regime $II'_l$, for $Pr<Ra^{-1}$.
\item Regime $IV'_l$ for $Ra^{-1}<Pr<1$.
\item Regime $IV'_u$ for $1<Pr<Ra^{1/3}$.
\item Regime $III'_\infty$ for $Ra^{1/3}<Pr<Ra^{2/3}$.
\end{itemize}
Table \ref{tab:ultimateregime} summarises the behaviour of the Nusselt, Reynolds, dissipation and the efficiency in these 4 regimes.
We see that in regime $IV'_l$ and $IV'_u$, $D_\epsilon$ never achieves saturation at infinite Rayleigh number, at variance with the zeroth law of turbulence, while efficiency saturates to constant independent of $Pr$. In contrast, in regime $II'_l$, dissipation tends to a constant, while efficiency goes to zero. Finally, in regime $III'_\infty$, the dissipation tends to zero and the efficiency to $\infty$.  Rayleigh-B\'enard convection therefore provides an example where efficiency and energy dissipation can have different behaviour s. Such difference could be used to offer an alternative detection of the ultimate regime.

\begin{table}
\begin{center}
\def~{\hphantom{0}}
\begin{tabular}
{ccccc}%
 Regime& $Nu$ &$Re$ &$D_\epsilon$ &${\cal E}$\\[3pt]
$II'_l$ &$Pr^{1/5} Ra^{1/5}$ &$Pr^{-3/5} Ra^{2/5}$ &$Pr^{0} Ra^{0}$&$Pr^{-1/5} Ra^{-1/5}$ \\
$III'_\infty$ &$Pr^{0} Ra^{1/3}$ &$Pr^{-1} Ra^{2/3}$ &$Pr^{1} Ra^{-2/3}$ &$Pr^{-1} Ra^{1/3}$ \\
$IV'_l$ &$Pr^{1/2} \frac{Ra^{1/2}}{(\log(Ra))^2}$ &$Pr^{-1/2} Ra^{1/2}$ &$Pr^{0} \frac{1}{(\log(Ra))^2}$ &$Pr^{0} Ra^{0}$ \\
$IV'_u$ &$Pr^{-1/2} \frac{Ra^{1/2}}{(\log(Ra))^2}$ &$Pr^{-1/2} Ra^{1/2}$ &$Pr^{-1} \frac{1}{(\log(Ra))^2}$ &$Pr^{0} Ra^{0}$ \\
\end{tabular}
\caption{Variation of the Nusselt number $Nu$, the Reynolds number $Re$, the dimensionless dissipation $D_\epsilon$ and the efficiency ${\cal E}$ as a function of the Rayleigh number $Ra$ and the Prandtl number $Pr$, in the four asymptotic ultimate regime identified by \cite{LS2024}. The two last columns are mere consequences of the definition of $D_\epsilon$ and ${\cal E}$. }
\label{tab:ultimateregime}
 \end{center}
\end{table}

\section{Discussion}

A recent work on the probabilistic representation of Kolmogorov's 1949 theory decomposes the turbulence phenomenology into three axioms. The first two being: (A1) The variance of any component of the velocity field remains finite and independent of the viscosity in the long-time and inviscid limit. (A2) The average energy dissipation remains finite in the inviscid limit. Clearly, the ``efficiency saturation'' discussed in the present study is connected to (A1), which ensures the ${\cal{E}}$ remains finite in the inviscid limit, so that $1/{\cal{E}}$ cannot tend to zero in the inviscid limit. In the same way, the axiom (A2) forbids the energy dissipation from diverging in the inviscid limit. However, none of these axioms ``guarantee'' the presence of convergence to a finite non-zero value in the inviscid limit. 

In view of the above discussion, it is interesting to note that the efficiency saturation is realised for the variety of flows that we consider here. Thus, the efficiency saturation is a robust property of turbulence. Also, the saturation is achieved following a power law, which in turn is associated with and motivated by the transitional behaviour of the solutions of the (formally) time-reversible Navier-Stokes equations. Our observations also suggest that the coefficients of the power law depend on the flow geometry: the behaviour is steeper (large exponent) for the pipe flows when compared to the homogeneous turbulence or the von K\'arm\'an flow. In the latter case, we found that the efficiency is fairly independent of the type of impellers that are used, even when they impart large anisotropies in kinetic energy. \

In the case of pipe flow, we show that saturation of the efficiency cannot be excluded but would be incompatible with the Prandtl law of the drag friction coefficient. In case the saturation of efficiency follows the power-law behaviour observed for homogeneous isotropic turbulence or von K\'arm\'an flow, it can explain the defect laws proposed by~\cite{Chen_Sreenivasan_2021} for the shear flows, both for the peak velocities and the energy dissipation. Indeed, because of the efficiency saturation, the peak velocity variances are necessarily bounded. Thus, by inverting the power law Eq.~\eqref{criticalLaw}, we obtain that the convergence of the kinetic energy (a proxy of the peak velocity variances) to its limit in accordance with a defect law $E\sim K-Re^{-\alpha}$, with an exponent $\alpha=1/3.5=0.28$. It is very close to the defect law with $\alpha=1/4$ proposed by~\cite{Chen_Sreenivasan_2021}. Note that imperfect agreement with the latter law has already been reported in ~\cite{Pirozzoli_2024}. They find a smaller exponent $\alpha=0.18$, but coincidentally, when extrapolating data to high Reynolds number, they obtain a saturation of dissipation rate to a value $0.28$. This saturation can also be explained in the present framework, as the defect law for kinetic energy also applies to the dimensionless energy dissipation $D_{\epsilon}$, provided $\epsilon$ obeys the dissipation anomaly. \

In the case of Taylor-Couette flow, the power law could not be firmly established, because more data would be needed.\

In the case of Rayleigh-B\'enard convection, we show that within the extended of the Grossmann-Lohse theory to the ultimate regime \citep{LS2024}, the efficiency saturates in the inviscid limit, while the dimensionless kinetic energy injection/dissipation goes to zero. This shows that efficiency and dimensionless energy dissipation are distinct quantities, that can vary independently of each other, provided the dimensionless energy dissipation remains bounded by the inefficiency.\

In summary, our study suggests that the efficiency saturation is an interesting empirical property of turbulence that may help in evaluating the ``closeness'' of experimental and numerical data to the true turbulent regime, wherein the kinetic energy saturates to its inviscid limit.

\section*{Funding}
This work received funding from the Ecole Polytechnique and from ANR BANG, grant agreement no. ANR-22-CE30-0025-01 and ANR TILT grant agreement no. ANR-20-CE30-0035. 

\section*{Acknowledgements}
We thank D. Lohse for providing us with the Taylor-Couette data, and O. Shishkina for providing us with the Rayleigh-B\'enard data. We thank Giorgio Krstulovic for providing data of reversible Navier-Stokes direct numerical simulations (case S2).

\bibliographystyle{unsrtnat}
\bibliography{efficiency_hal}

\end{document}